\pdfoutput=1

\documentclass[11pt,twoside,a4paper,cmspaper,final,collab]{cms-tdr}
\def\svnVersion{423863P}\def\cmsCernNoTag{CERN-EP-2016-198}\def\cmsCernDate{\today}\def\cmsMessage{Published in Physics Letters B as \href{http://dx.doi.org/10.1016/j.physletb.2017.05.074}{\doi{10.1016/j.physletb.2017.05.074.}}}

\begin{document}\cmsNoteHeader{BPH-15-004}

\hyphenation{had-ron-i-za-tion}
\hyphenation{cal-or-i-me-ter}
\hyphenation{de-vices}
\RCS$Revision: 405687 $
\RCS$HeadURL: svn+ssh://svn.cern.ch/reps/tdr2/papers/BPH-15-004/trunk/BPH-15-004.tex $
\RCS$Id: BPH-15-004.tex 405687 2017-05-20 10:38:46Z alverson $
\newlength\cmsFigWidth
\ifthenelse{\boolean{cms@external}}{\setlength\cmsFigWidth{0.90\columnwidth}}{\setlength\cmsFigWidth{0.48\textwidth}}
\ifthenelse{\boolean{cms@external}}{\providecommand{\cmsLeft}{top\xspace}}{\providecommand{\cmsLeft}{left\xspace}}
\ifthenelse{\boolean{cms@external}}{\providecommand{\cmsRight}{bottom\xspace}}{\providecommand{\cmsRight}{right\xspace}}
\providecommand{\mub}{\ensuremath{\mu\mathrm{b}}\xspace}
\cmsNoteHeader{BPH-15-004}
\newcommand{\x}{\ensuremath{\phantom{0}}}
\newcommand{\xx}{\ensuremath{\phantom{00}}}
\title{Measurement of the differential inclusive $\PBp$ hadron cross sections in pp collisions at $\sqrt{s}=13\TeV$}
\date{\today}
\abstract{
The differential cross sections for inclusive production of $\PBp$ hadrons are measured
as a function of the $\PBp$ transverse momentum $\pt^\PB$ and rapidity $y^\PB$ in pp collisions at a centre-of-mass energy of 13\TeV,
using data collected by the CMS experiment that correspond to an integrated luminosity of
48.1\pbinv. The measurement uses the exclusive decay channel
$\PBp \to \PJGy \PKp$, with $\PJGy$ mesons that decay to a pair of muons.
The results show a reasonable agreement with theoretical calculations within the uncertainties.
}
\hypersetup{%
pdfauthor={CMS Collaboration},%
pdftitle={Measurement of the total and differential inclusive B+ hadron cross sections in pp collisions at sqrt(s)=13 TeV},%
pdfsubject={CMS},%
pdfkeywords={CMS, physics, B-physics}}
\maketitle
\section{Introduction}
\label{sec:introduction}
Measuring the production of hadrons that contain b quarks plays an important role in testing quantum chromodynamics (QCD).
Such studies have been carried out by several experiments,
including UA1~\cite{UA1-87,UA188} at CERN,
as well as CDF~\cite{TeVI-CDF3, TeVI-CDF6,TeVII-CDF1, TeVII-CDF2} and \DZERO~\cite{TeVI-D01, TeVI-D04} at Fermilab.
The most recent measurements are from the
ATLAS~\cite{Aad:2012jga, ATLAS:2013cia},
CMS~\cite{Khachatryan:2011mk, Khachatryan:2011hf, Chatrchyan:2011pw, Chatrchyan:2011vh, Chatrchyan:2012xg, Chatrchyan:2012hw, Khachatryan:2014nfa}, and LHCb~\cite{Aaij:2012jd, Aaij:2013noa, Aaij:2014ija}
Collaborations at the CERN LHC
in $\Pp\Pp$ collisions at centre-of-mass energies at 7 and 8\TeV.
Similar studies at the higher LHC energy of 13\TeV provide a
new test of theoretical calculations~\cite{Aaij:2015rla, Aaij:2016avz}.

This Letter describes a measurement of the inclusive $\PBp$ differential production cross sections
as a function of the transverse momentum ($\pt^{\PB}$) and rapidity ($y^{\PB}$) of the $\PBp$ meson
(charge conjugation is implied throughout this paper).
The analysis is based on data collected at the LHC with 50\unit{ns} bunch spacings by the CMS experiment at $\sqrt{s}=13\TeV$ that
correspond to an integrated luminosity of 48.1\pbinv.
The measurement is based on the inclusive channel $\Pp\Pp \to \PBp \mathrm{X} \to \PJGy \PKp \mathrm{X}$,
with the $\PJGy$ mesons decaying into a pair of muons.
The measured cross sections are compared to
\PYTHIA~\cite{Sjostrand:2007gs} and FONLL~\cite{Cacciari:1998it,Cacciari:2001td} calculations.
\section{The CMS detector and trigger}
The central feature of the CMS apparatus is a superconducting solenoid of 6\unit{m} internal diameter,
providing a magnetic field of 3.8\unit{T}. A silicon pixel and strip tracker,
a lead tungstate crystal electromagnetic calorimeter, and a brass and scintillator hadron calorimeter reside within the magnetic volume of the solenoid.
The inner tracker measures charged particles within the pseudorapidity range $\abs{\eta}< 2.5$.
Muons are measured with detectors made using three technologies:
drift tubes, cathode strip chambers, and resistive-plate chambers.
Matching stubs in the muon system to tracks measured in the silicon tracker result in a transverse momentum (\pt) resolution
better than 1.5\% for a typical muon in this analysis~\cite{Chatrchyan:2012xi}.
A more detailed description of the CMS detector, together with a definition of the coordinate system and kinematic variables, can be found in Ref.~\cite{:2008zzk}.

The triggers have two levels of activation:
the first-level (L1) trigger is based on the information provided by the muon detectors,
while the ``high-level trigger" (HLT) uses information from the silicon tracker to filter the events.
Two L1 trigger requirements are used:
one requires two muons in the barrel region ($\abs{\eta} < 1.6$),
without explicitly imposing a minimum \pt value;
the other accepts two muons with relaxed pseudorapidity restrictions
(i.e. $\abs{\eta} < 2.4$) but requires at least one muon to have $\pt > 10\GeV$.
The HLT requires the two muons to be of opposite charge,
to lie within $\abs{\eta} < 2.4$, and to have $\pt > 4\GeV$.
The dimuon invariant mass must be in the range 2.9--3.3\GeV, and
the $\chi^2$ probability of the dimuon fit (imposing a common vertex)
must be greater than 10\%.
Furthermore, the signal purity from the trigger is enhanced by requiring the distance between the
dimuon vertex and the interaction point (the mean $\Pp\Pp$ collision position or beam spot, which is determined for each set of events collected during a period of 23 seconds) in the transverse plane be larger than three times its uncertainty;
this requirement preferentially selects dimuons from ``nonprompt" $\PJGy$ meson decays, while rejecting almost all
promptly produced $\PJGy$ mesons.
Also, the $\PJGy$ meson momentum vector reconstructed at the HLT stage must point back to the
interaction point in the transverse plane. This condition is imposed by requiring $\cos(\alpha) > 0.9$, where $\alpha$ is the angle between the $\PJGy$
momentum vector and
the vector pointing from the interaction point to the dimuon vertex.
Finally, the $\PJGy$ candidate is combined with any other charged particle of $\pt > 0.8\GeV$,
and the three-track fit to a common secondary vertex is performed.
The HLT requires at least one of the fits to have a $\chi^{2}$ per degree of freedom smaller than 10.
\section{Event reconstruction and selection}
\label{sec:EvtSel}
The first step in the reconstruction of the $\PBp\to\PJGy\PKp$ decays is the selection of
events containing a pair of muons originating
from the decay of a $\PJGy$ meson.
The muons are required to have at least one reconstructed segment in the muon detectors that matches
the extrapolated position of a track reconstructed in the silicon tracker, to satisfy
$\pt > 4.2\GeV$, $\abs{\eta}<2.1$, and to have good quality in the fit to a track.
The muon tracks are required to intersect a cylinder of 0.3\unit{cm} radius in the transverse plane
and 20\unit{cm} length along the beam line relative to the interaction point.

Candidate $\PJGy$ mesons are reconstructed by combining pairs of oppositely charged muons
having an invariant mass within $\pm$150\MeV of the nominal $\PJGy$ meson mass~\cite{Agashe:2014kda}.
Each $\PJGy$ candidate must have $\pt > 8\GeV$, and a
$\chi^2$ probability for a fit to the dimuon vertex larger than 10\%.
Both muons must be either within $\abs{\eta} < 1.6$ or one of the muons must have $\pt > 11\GeV$.
Candidate $\PBp$ mesons are reconstructed
by combining a $\PJGy$ candidate with each charged track in the event having $\pt > 1\GeV$.
A kaon mass hypothesis is assumed for the tracks
and the $\chi^{2}$ per degree of freedom of the track fit is required to be less than 5.
A kinematic fit is performed to the dimuon-track combination, constraining the dimuon mass to the nominal $\PJGy$ mass.
The three-track combination must be compatible with having a common vertex with a
$\chi^{2}$ probability larger than 10\%
and a reconstructed invariant mass, $M_{\PJGy \PK}$, in the range 5--6\GeV.
The significance in the transverse decay length,
defined as the distance between the $\mu\mu\PK$ vertex and
the interaction point in the transverse plane, divided by its uncertainty,
is required to exceed~3.5.
Also, the cosine of the angle between the $\PBp$ candidate momentum and the vector
pointing from the interaction point to the $\mu\mu\PK$ vertex
in the transverse plane must be greater than 0.99.
Most of the selected $\PBp$ candidates have a transverse decay length greater than 300\micron.
Only a small fraction (${<}1\%$) of events contain two reconstructed $\PBp$ candidates; all reconstructed candidates are included in the analysis.
This analysis is insensitive to the number of proton-proton interactions occurring in the same or nearby bunch crossings.

The combinatorial background
arises from the spurious combination of a promptly produced $\PJGy$ meson
or a $\PJGy$ meson from a $\PB$ hadron decay with an uncorrelated charged particle.
The former case is suppressed by the requirement on the reconstructed decay length of the $\PBp$ candidate.
Given the excellent dimuon mass resolution at the $\PJGy$ mass
and the good muon identification performance of the CMS detector,
the background level under the $\PJGy$ peak is very small.
Other backgrounds arise from misreconstructed $\cPqb$ hadron decays,
such as $\PB \to \PJGy + \text{hadrons}$ (including, e.g. $\PJGy \PKst$),
which contribute a broad structure in the mass region $M_{\PJGy \PK} < 5.15\GeV$.
Additional background from
Cabibbo--Kobayashi--Maskawa-suppressed $\PBp \to \PJGy \Pgpp$ decays with a mass misassignment to the pion track forms
a tiny excess just above the $\PJGy \PKp$  signal.
\section{Reconstruction efficiency and acceptance}
\label{sec:EffAcc}
The detection and trigger efficiencies and the geometrical acceptance
are evaluated through Monte Carlo simulation studies using large
samples of signal events generated in \PYTHIA~8.205 (using the CUETP8M1
tune~\cite{Khachatryan:2015pea} and the NNPDF2.3 parton distribution functions~\cite{Ball:2013hta}) and
processed by the simulation framework of the CMS detector based on
\GEANTfour~\cite{GEANT4:2003}. The decays $\PBp \to \PJGy \PKp$ are modelled with the SVS model
of the \EVTGEN~1.3.0~\cite{Lange:2001} generator.
The product of the efficiency and acceptance is defined as the fraction of simulated $\PBp \to \PJGy \PKp \to \mu^{+} \mu^{-} \PKp$ decays,
generated in the phase-space region
of $10 \le \pt^{\PB} < 17 \GeV$ and  $\abs{y^{\PB}} < 1.45$,  and in the region of $17 \le \pt^{\PB} < 100 \GeV$ and  $\abs{y^{\PB}} < 2.1$,
that survive the selection criteria.
These values range from $0.8\%$ for $\pt^{\PB} \approx 10\GeV$ to
$20\%$ for $70 < \pt^{\PB} < 100\GeV$, and  from
$3.6\%$ for $\abs{y^{\PB}} \approx 0$ to
$2.5\%$ for $1.8 < \abs{y^{\PB}} < 2.1$.

The trigger and muon reconstruction efficiencies are also measured from a data sample consisting
of inclusive $\PJGy \to \mu^{+} \mu^{-}$ decays, using a technique similar to
that described in Ref.~\cite{Khachatryan:2010yr}, where one muon is identified with stringent quality
requirements and the second muon is identified using information either
from the tracker (to measure the trigger and muon identification efficiencies) or from
the external muon system (to measure the silicon tracker efficiency).
These efficiencies are compared to those from simulation studies,
in bins of muon $\pt$ and $\eta$, and are found to  agree within their uncertainties.
The measured efficiencies of the track reconstruction and of the vertex-quality requirement
are also found to be compatible with the simulations.
\section{Extraction of the signal}
The $\PBp$ differential cross section $\rd\sigma/\rd\pt^{\PB}$ is measured in 2 bins of $\pt^{\PB}$ between 10 and 17\GeV in a restricted $y^{\PB}$ range ($\abs{y^{\PB}}<1.45$) and in 7 bins of $\pt^{\PB}$ in the extended $y^{\PB}$ range ($\abs{y^{\PB}}<2.1$) for $\pt^{\PB}$ between 17 and 100\GeV. The corresponding differential cross section $\rd\sigma/\rd y^{\PB}$ is measured in 6 (2) bins of $\abs{y^{\PB}}$ for $\pt^{\PB}$ between 10--100\GeV (17--100\GeV).
The signal yield is extracted with an extended unbinned maximum-likelihood fit to the invariant mass distribution of the
$\PBp$ candidates in each of the $\pt^{\PB}$ or $\abs{y^{\PB}}$ bins.
The signal component is modelled by the sum of two Gaussian functions (representing the ``core" and a ``tail").
The relative mean of the tail Gaussian function is fixed with respect to that of the core Gaussian function,
following the shapes obtained from simulated samples.
\begin{figure}[t]
\centering
\includegraphics[width=\cmsFigWidth]{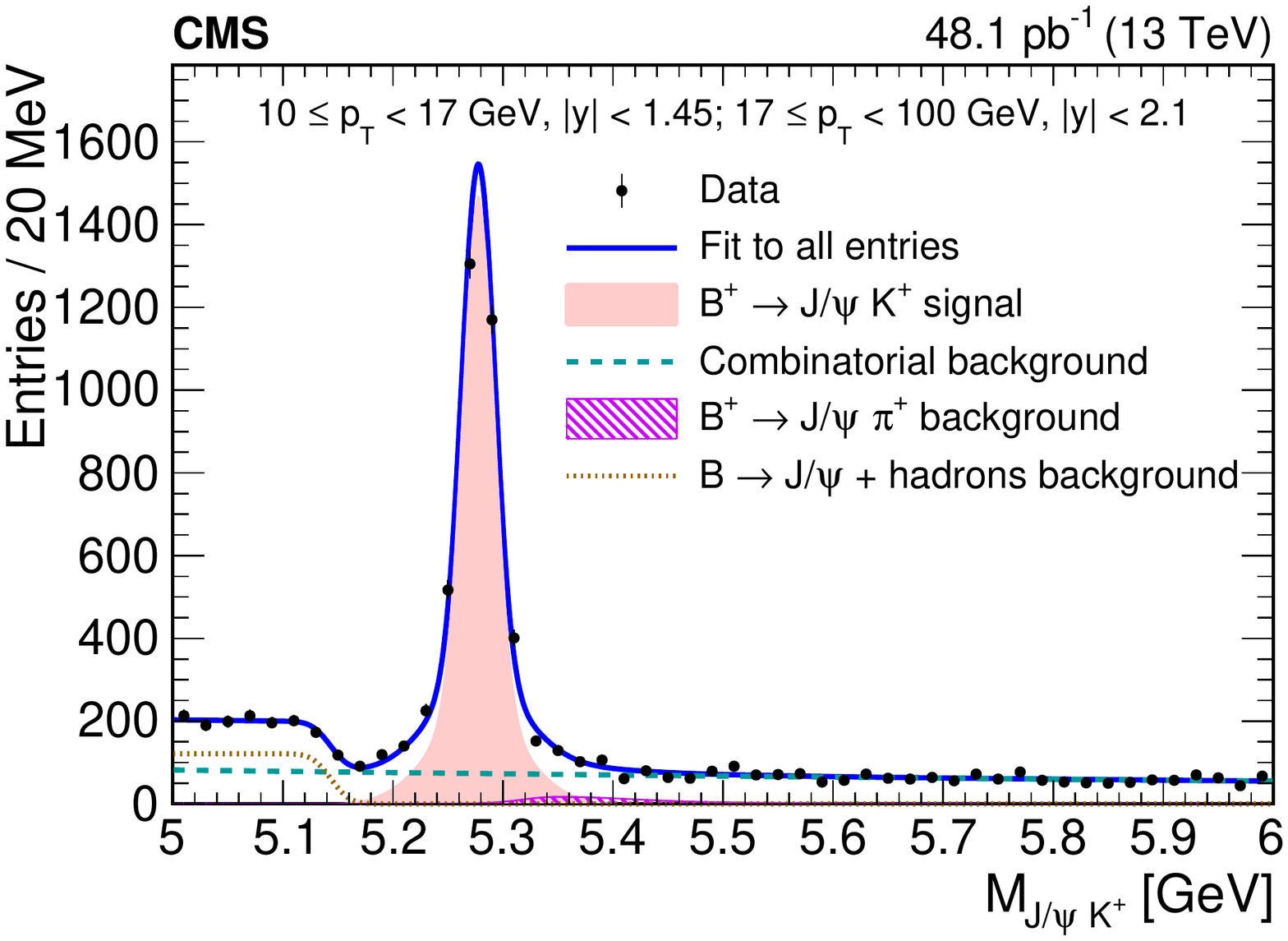}
\caption{
Invariant mass distribution of $\PBp \to \PJGy \PKp$ candidates,
integrated over the phase-space region of
$10 \le \pt^{\PB} < 17 \GeV$ and $\abs{y^{\PB}} < 1.45$,  and of  $17 \le \pt^{\PB} < 100 \GeV$ and  $\abs{y^{\PB}} < 2.1$.
The solid curve shows the result of the fit.
The shaded and hatched areas represent, respectively, the  $\PJGy \PKp$ signal and the $\PJGy \Pgpp$ component,
while the dashed and dotted curves represent the
combinatorial and misreconstructed $\PB \to \PJGy + \text{hadrons}$ backgrounds, respectively.}
\label{fig:invMass10_100}
\end{figure}
\begin{figure}[tb]
\includegraphics[width=\cmsFigWidth]{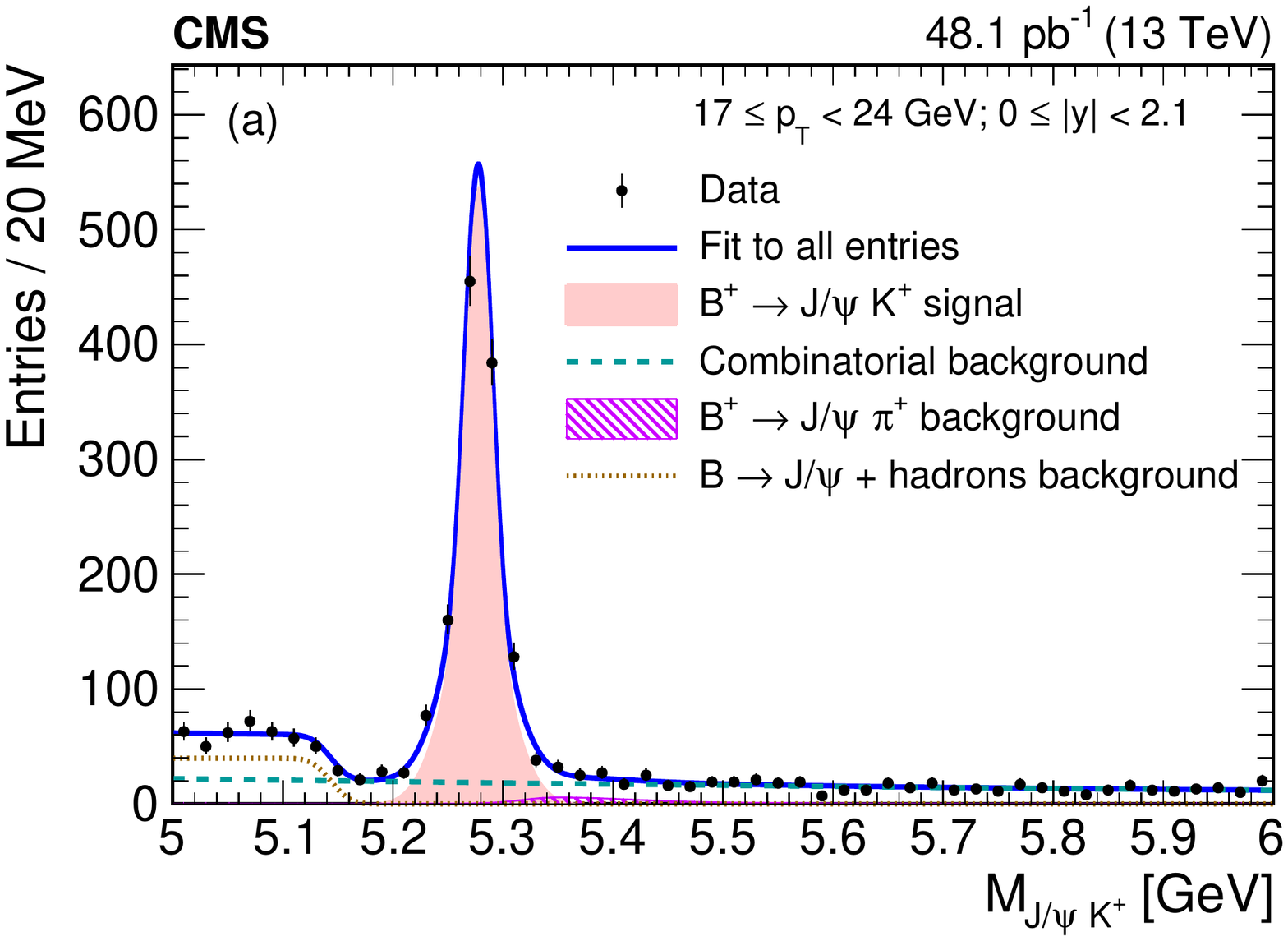}
\includegraphics[width=\cmsFigWidth]{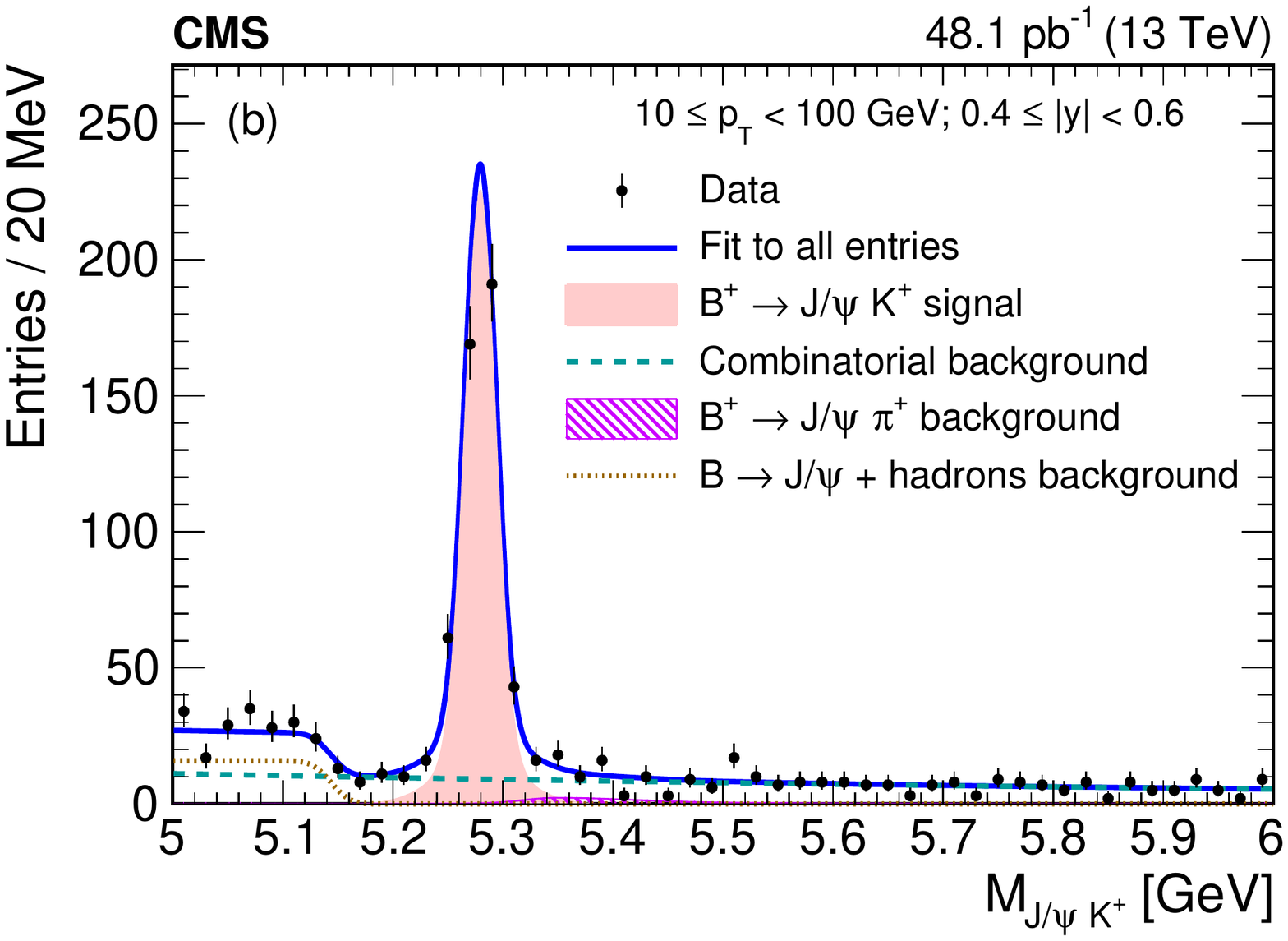}
\caption{
Invariant mass distributions of the $\PBp \to \PJGy \PKp$ candidates in the regions of
(a) $17 \le \pt^{\PB} < 24 \GeV$, $\abs{y^{\PB}} < 2.1$, and
(b) $10 \le \pt^{\PB} < 100 \GeV$, $0.4 \le \abs{y^{\PB}} < 0.6$.
The solid curve shows the result of the fit.
The shaded (hatched) area represents the signal ($\PBp \to \PJGy \Pgpp$) component,
while the dashed and dotted curves represent the
combinatorial and misreconstructed $\PB \to \PJGy + \text{hadrons}$ background components, respectively.}
\label{fig:invMassSub}
\end{figure}
The relative normalization and width of the tail Gaussian function are also fixed to the parameters obtained from simulated samples
for the $\pt^{\PB}$ bins above 50 GeV and $\abs{y^{\PB}}$ bins above 1.45, to account for the limited size of the data sample.
The combinatorial background from the inclusive $\PJGy$ meson production is modelled by an exponential function.
The background from misreconstructed $\PB \to \PJGy + \text{hadrons}$ decays is represented by an error function.
The normalization of misreconstructed $\PB \to \PJGy + \text{hadrons}$ decays relative to the signal is determined from a fit to all selected $\PBp$ candidates,
and is fixed to this value in the fits that are performed in individual $\pt^{\PB}$ or $\abs{y^{\PB}}$ bins.
The contribution from the decay $\PBp \to \PJGy \Pgpp$ is modelled through a sum of three Gaussian functions;
the relative yield of the $\PJGy \Pgpp$ to the signal $\PJGy \PKp$  is fixed by their decay branching fractions~\cite{Agashe:2014kda}.
Figure~\ref{fig:invMass10_100} shows the invariant mass distribution
of all the $\PBp$ candidates,
compared to the corresponding sum of the signal and background distributions obtained from the maximum-likelihood fit. Typical invariant mass distributions of the $\PBp$ candidates
in one of the $\pt^{\PB}$ bins and in one of the $\abs{y^{\PB}}$ bins
are shown in Fig.~\ref{fig:invMassSub}.
The $\PBp \to \PJGy \Pgpp$ background, shown by the hatched area, is centered around 5.4\GeV, but is so small as to be almost invisible in the figures.
The dip in the measured invariant mass distributions around 5.17\GeV is caused by the shape of the $\PB \to \PJGy + \text{hadrons}$ background distribution, which falls abruptly in that region.
\section{Systematic uncertainties}
\label{sec:SysErr}
The measured cross section is affected by systematic uncertainties in the extraction of signal,
efficiencies, branching fractions, and integrated luminosity, as summarized in Table~\ref{tab:systematics}.
The dominant effects are associated with the models used in the likelihood fits,  the $\PBp$ kinematic distributions, and
the estimation of the muon identification and reconstruction.
The total uncertainty is evaluated as the sum in quadrature of the individual contributions.

The uncertainty associated with
the trigger criteria is evaluated by comparing the trigger efficiencies in data and
simulations for an event sample recorded using an inclusive $\PJGy$ trigger with higher-$\pt$ thresholds.
The muon identification and reconstruction performances are studied using a large sample of
inclusive $\PJGy \to \mu^+\mu^-$ events. The efficiencies in data and simulated events are found to be consistent,
and residual differences are considered as systematic uncertainties.
The uncertainty associated with the
alignment of the detector is examined by
comparing events simulated with different detector conditions, and assigning an uncertainty of 2.8\%.
Through a comparison of $\chi^2$ distributions in data and simulations,
the uncertainty in the $\PBp$ vertex reconstruction efficiency is estimated to be 1.4\%.
In addition, the uncertainty coming from the finite size of the simulated samples and a
systematic uncertainty of 3.9\% in the charged-particle track reconstruction efficiency~\cite{TRK-11-001} are also taken into account.
The integrated luminosity is measured with an uncertainty of 2.3\%~\cite{LUM-15-001}, while the uncertainty associated
with the $\PBp \to \PJGy \PKp \to \mu^+\mu^- \PKp$ branching fraction is 3.1\%~\cite{Agashe:2014kda}.

The systematic uncertainty associated with the
modeling of the signal shape is evaluated by changing the model to the sum of
three Gaussian functions, or a Gaussian function plus a Crystal Ball function~\cite{Skwarnicki:1986xj}.
The uncertainty from the model of the combinatorial background
is evaluated by changing the exponential function to a second-order polynomial.
The systematic uncertainty associated with the modeling of the mass distribution of
misreconstructed $\PB \to \PJGy + \text{hadrons}$ events is evaluated
by shifting the mass-threshold parameter in the error function by $\pm10\MeV$.
The uncertainty associated with the $\PBp \to \PJGy \Pgpp$ component is estimated by changing its branching fraction
by its uncertainty~\cite{Agashe:2014kda},
and by shifting its mass value by $\pm15\MeV$ in the likelihood fits.
Systematic uncertainties owing to the finite resolution of the reconstructed $\pt^{\PB}$ and $y^{\PB}$ are determined by
examining the generator information in the simulated samples;
half of the bin-to-bin-migrated events are taken as the corresponding uncertainty.
The uncertainties associated with the $\pt^{\PB}$ and $y^{\PB}$ distributions in the generation of
simulated events are evaluated with event-by-event weights determined from the differences between
the distributions in \PYTHIA and the FONLL calculations. The latter uses
a fixed-order perturbative QCD approach, with a next-to-leading-logarithm approximation, and the NNPDF3.0 parton distribution functions~\cite{Ball:2014uwa}.
The uncertainty associated with the parton distribution functions is found to be less than 0.7\%, which is estimated using the PDF4LHC prescription~\cite{Botje:2011sn, Alekhin:2011sk} with the uncertainty sets provided by the NNPDF2.3~\cite{Ball:2013hta}.
The effect of the systematic uncertainty in the $\PBp$ lifetime (0.3\%) is also included.
\begin{table}[tb]
\centering
\topcaption{Summary of the relative systematic uncertainties in the measured $\PBp$ production cross sections.
The ranges given  reflect the uncertainties over the $\pt^{\PB}$ and $y^{\PB}$ bins.}
\setlength\tabcolsep{3pt}%
\newlength\ghmtableone
\ifthenelse{\boolean{cms@external}}{\setlength\ghmtableone{\columnwidth}}{\setlength\ghmtableone{0.65\columnwidth}}
\resizebox{\ghmtableone}{!}{
\begin{tabular}[h]{lc}
\hline
\multirow{2}{*}{Systematic sources} & Relative \\
 & uncertainties (\%) \\
\hline
Muon trigger, identification,  &  \multirow{2}{*}{\x6.0--14} \\
~~~~and reconstruction &   \\
Detector alignment & 2.8 \\
$\PBp$ vertex reconstruction & 1.4 \\
Size of simulated samples & 0.5--3.9 \\
Track reconstruction efficiency & 3.9 \\
$\PBp \to \PJGy (\to \mu^+\mu^-) \PKp$ branching fraction & 3.1 \\
Model in likelihood fits & 1.0--6.4 \\
Bin-to-bin migration & 0.4--3.7 \\
$\PBp$ kinematic distributions & \x0.4--11 \\
Parton distribution functions & 0.1--0.7 \\
$\PBp$ lifetime & 0.3 \\
\hline
Total (excluding the integrated luminosity) & \x9.1--16 \\
Integrated luminosity & 2.3 \\
\hline
\end{tabular}
}
\label{tab:systematics}
\end{table}
\section{Results}
\label{sec:results}
The differential cross sections for $\PBp$ production as a function of $\pt^{\PB}$, for $\abs{y^{\PB}} < 1.45$, or for $\abs{y^{\PB}} < 2.1$,
$\rd\sigma/\rd\pt^{\PB}$, and as a function of $\abs{y^{\PB}}$ (averaged for positive and negative rapidity)
for $10<\pt^{\PB}<100\GeV$, or for $17<\pt^{\PB}<100\GeV$, $\rd\sigma/\rd y^{\PB}$, are defined as
\begin{equation}
\begin{split}
\frac{{\rd}\sigma(\Pp\Pp \rightarrow \PBp \mathrm{X})}{{\rd}\pt^{\PB}} &= \frac{n_\text{sig}(\pt^{\PB})}{2\,A(\pt^{\PB})\,\epsilon(\pt^{\PB})\,\mathcal{B}\,\mathcal{L}\,\Delta \pt^{\PB}}, \\
\frac{{\rd}\sigma(\Pp\Pp \rightarrow \PBp \mathrm{X})}{{\rd}y^{\PB}} &= \frac{n_\text{sig}(\abs{y^{\PB}})}{2\,A(\abs{y^{\PB}})\,\epsilon(\abs{y^{\PB}})\,\mathcal{B}\,\mathcal{L}\,\Delta y^{\PB}},
\end{split}
\label{Eq:crosssectionFor}
\end{equation}
where $n_\text{sig}(\pt^{\PB})$ and $n_\text{sig}(\abs{y^{\PB}})$ are the signal yields in the
$\pt^{\PB}$ or $\abs{y^{\PB}}$ bins, obtained in the maximum-likelihood fits; and
$\Delta \pt^{\PB}$ and $\Delta y^{\PB}$ = $2\Delta\abs{y^{\PB}}$ are the corresponding bin widths;
$\mathcal{L}$ is the integrated luminosity.
The total branching fraction $\mathcal{B}$ is the product of the individual
$\mathcal{B}(\PBp \to \PJGy \PKp) = (1.026 \pm 0.031) \times 10^{-3}$ and
$\mathcal{B}(\PJGy \to \mu^{+}\mu^{-}) = (5.961 \pm 0.033) \times 10^{-2}$ values~\cite{Agashe:2014kda}.
The factor of two in the denominator reflects the choice used to quote the cross section
for a single charge (taken to be the $\PBp$), where $n_\text{sig}$ includes both charge states.
Efficiencies and signal yields for $\PBp$ and $\PBm$ are found to be compatible within uncertainties.
The products of efficiency and acceptance, $A(\pt^{\PB})\,\epsilon(\pt^{\PB})$ and $A(\abs{y^{\PB}})\,\epsilon(\abs{y^{\PB}})$,
are calculated for each bin.

\begin{figure*}[tbph]
\centering
\includegraphics[width=\cmsFigWidth]{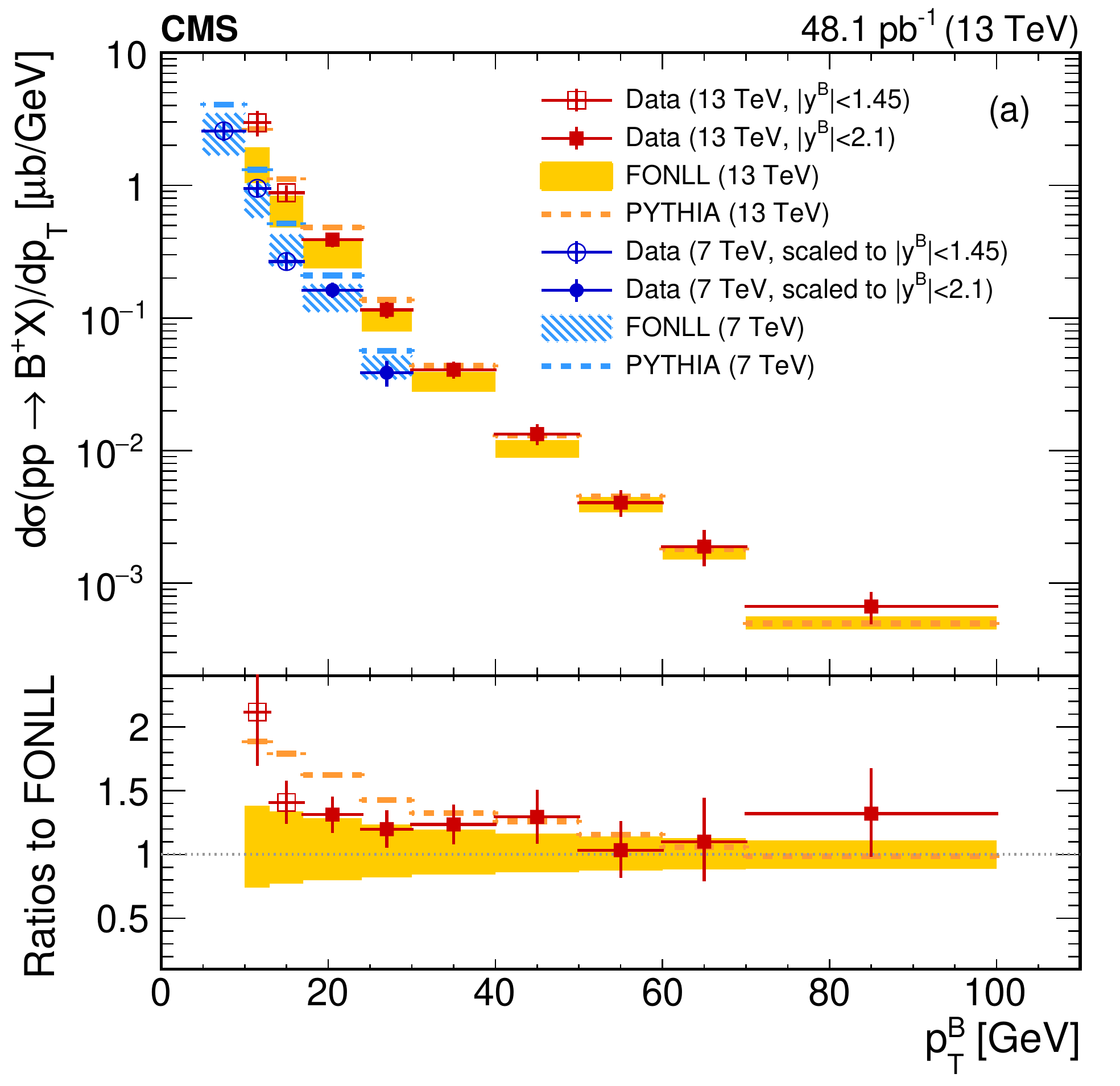}\hspace{2em}
\includegraphics[width=\cmsFigWidth]{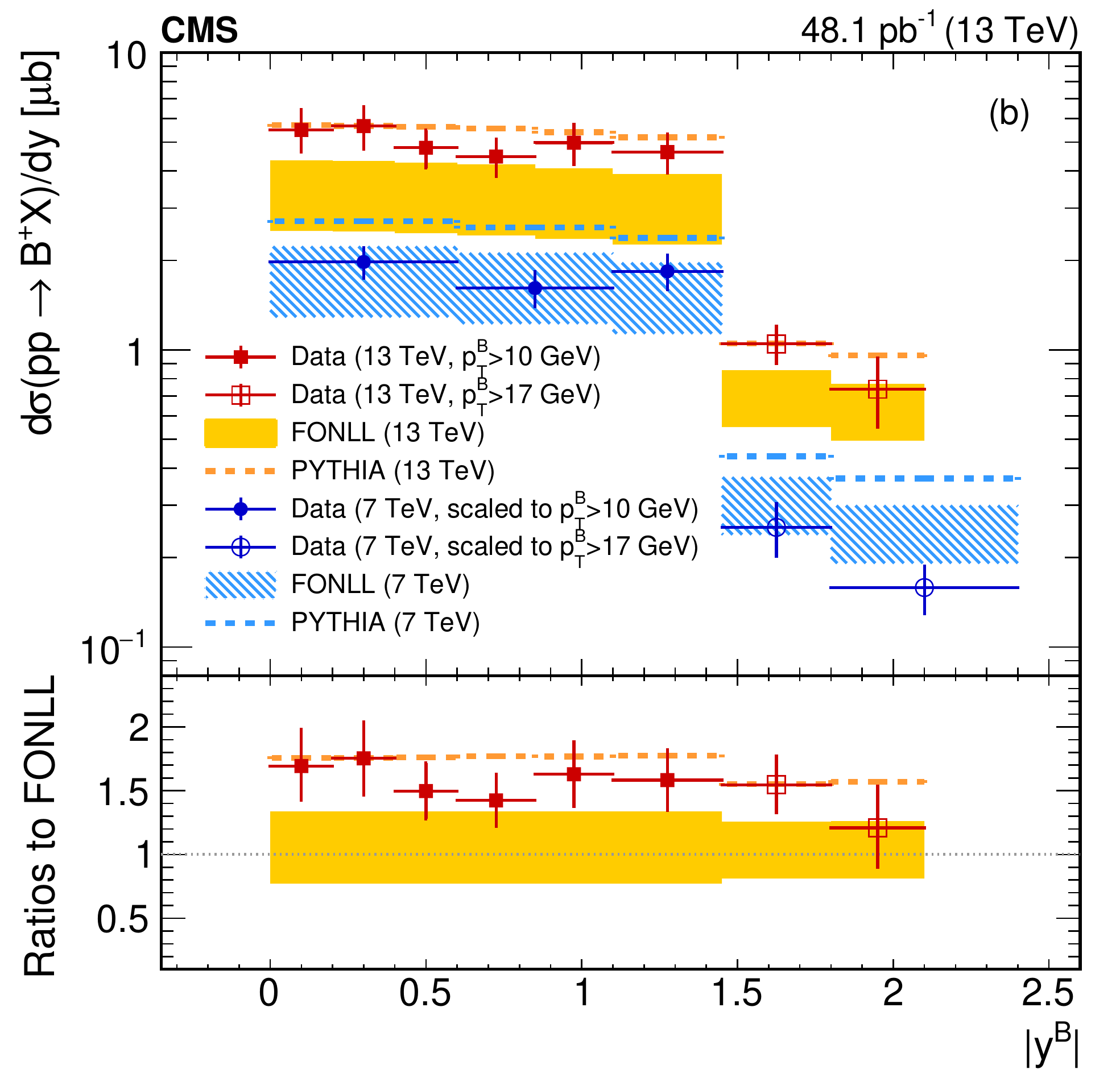}
\caption{$\PBp$ differential production cross sections (a) $\rd\sigma/\rd\pt^{\PB}$ for $\abs{y^{\PB}} < 1.45$ or $\abs{y^{\PB}} < 2.1$,
and (b) $\rd\sigma/\rd y^{\PB}$ for $10 < \pt^{\PB} <100 \GeV$ or $17 < \pt^{\PB} <100 \GeV$, at $\sqrt{s} = 13\TeV$
(squares, this measurement).
The previous CMS measurements from $\sqrt{s}=7\TeV$ data~\cite{Khachatryan:2011mk} (circles) are also presented as a function of $\pt^{\PB}$ ($y^{\PB}$), scaled to
the phase-space region of $\abs{y^{\PB}} < 2.1$ or $\abs{y^{\PB}} < 1.45$ ($10 < \pt^{\PB} <100 \GeV$ or
$17 < \pt^{\PB} <100 \GeV$).
The vertical bars show the total uncertainty in the measured cross sections, and the horizontal bars represent the bin width.
The calculations from FONLL
and \PYTHIA are shown as shaded boxes and dashed lines, respectively.
The bottom panels display the ratio of the data at 13\TeV to the FONLL predictions (points) and
the ratios of the \PYTHIA to the FONLL calculations (dashed lines), with the shaded region displaying the
uncertainties in the FONLL predictions.
}
\label{fig:crosssection}
\end{figure*}

\begin{figure*}[tbph]
\centering
\includegraphics[width=\cmsFigWidth]{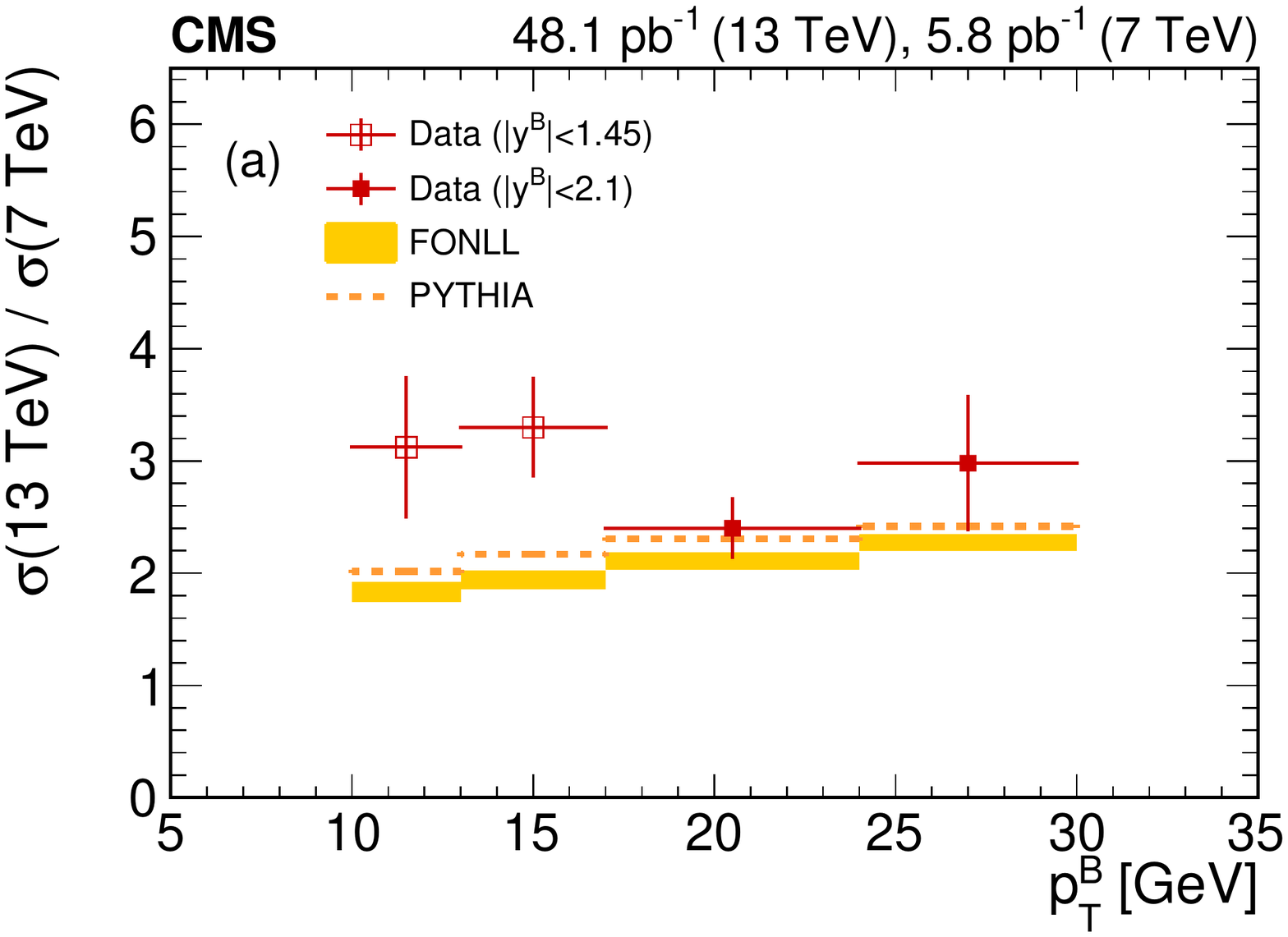}
\includegraphics[width=\cmsFigWidth]{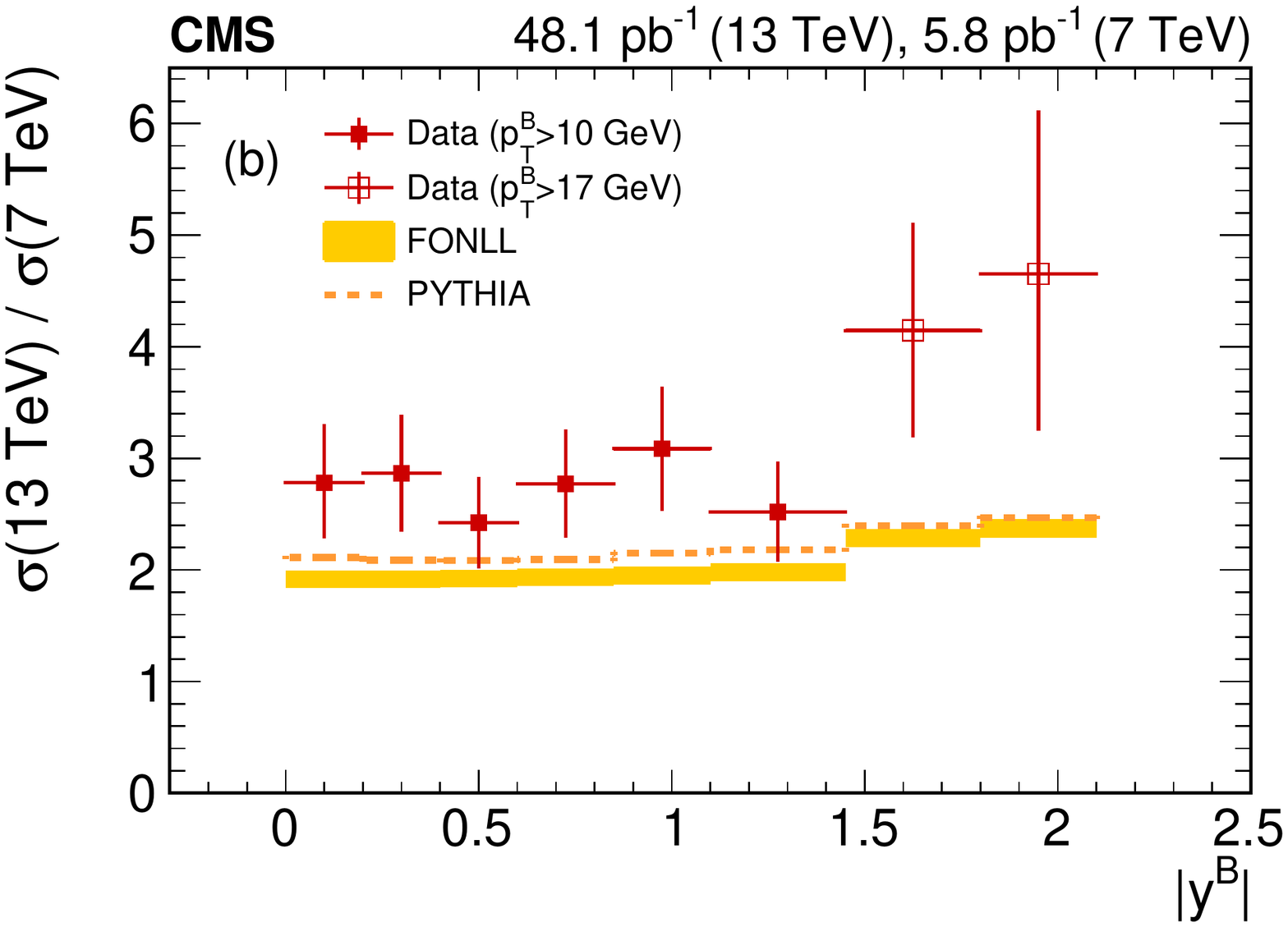}
\caption{Ratios of $\PBp$ differential production cross sections at $\sqrt{s}=13\TeV$ and at $\sqrt{s}=7\TeV$
as (a) a function of $\pt^{\PB}$ for $\abs{y^{\PB}} < 1.45$ or $\abs{y^{\PB}} < 2.1$ and (b) as a function of $\abs{y^{\PB}}$
for $10 < \pt^{\PB} <100 \GeV$ or $17 < \pt^{\PB} <100 \GeV$.
The vertical bars show the total uncertainty in the measured ratios of the cross sections, and the horizontal bars represent the bin width.
The calculations from FONLL
and \PYTHIA are shown as shaded boxes and dashed lines, respectively.
}
\label{fig:crosssection-ratio}
\end{figure*}

Table~\ref{tab:crosssection} summarizes the event yields, efficiencies, and the differential cross sections for the various $\pt^{\PB}$ and $y^{\PB}$ bins.
The differential cross sections as a function of $\pt^{\PB}$, integrated over $\abs{y^{\PB}} < 1.45$ or over $\abs{y^{\PB}} < 2.1$,
and as a function of $y^{\PB}$, integrated over $10 < \pt^{\PB} <100 \GeV$ or over $17 < \pt^{\PB} <100 \GeV$, are shown in
Fig.~\ref{fig:crosssection} (a) and (b), respectively, where they are compared to
FONLL~\cite{Cacciari:1998it,Cacciari:2001td} (shaded boxes) and \PYTHIA (dashed lines) calculations.
The uncertainties in the FONLL calculations include the effects from the renormalization and factorization scales, the mass of the bottom quark,
and the uncertainties in the parton distribution functions, which are calculated according to the NNPDF3.0 uncertainty sets~\cite{Ball:2014uwa}.
The bottom panels display the ratio of the data to the FONLL predictions;
the ratios of the \PYTHIA to the FONLL calculations are shown as dashed lines.
The previous CMS measurements from $\sqrt{s}=7\TeV$ data~\cite{Khachatryan:2011mk} are presented as a function of $\pt^{\PB}$, scaled to
the phase-space region of $\abs{y^{\PB}} < 2.1$ or $\abs{y^{\PB}} < 1.45$, and as a function of $y^{\PB}$, scaled to $10 < \pt^{\PB} <100 \GeV$ or
$17 < \pt^{\PB} <100 \GeV$.
The extrapolations are carried out using the kinematic distributions from generated \PYTHIA events, and
an additional systematic uncertainty is included based on a comparison of
extrapolations obtained with \PYTHIA to those obtained with FONLL.
Measurements are in good agreement with the theoretical predictions of both FONLL and \PYTHIA at high $\pt^{\PB}$,
while, at low $\pt^{\PB}$, the measurements tend to favour a higher cross section than estimated by FONLL and smaller than estimated by \PYTHIA.
The differential cross section as a function of $\abs{y^{\PB}}$ is in agreement with both predictions, within the uncertainties.
The ratios of the differential cross section measurements at $\sqrt{s}=13\TeV$ and $\sqrt{s}=7\TeV$, as well as
the FONLL and \PYTHIA calculations, are shown in Fig.~\ref{fig:crosssection-ratio}. The correlated uncertainties, including muon identification, decay branching fractions, tracking and vertexing, cancel out or are reduced in the evaluations of the ratios, and the measurements prefer higher values compared to the predictions along both $\pt^\PB$ and $\abs{y^\PB}$.
\begin{table*}[t]
\centering
\topcaption{The ranges in $\pt^{\PB}$ and $\abs{y^{\PB}}$, signal yields $n_\text{sig}$, acceptance times efficiency $A\epsilon$, and
measured differential cross sections ${{\rd}\sigma}/{{\rd}\pt^{\PB}}$ and ${{\rd}\sigma}/{{\rd}y^{\PB}}$, compared to the FONLL
and \PYTHIA predictions.
The three uncertainties in the measured cross sections refer to the statistical, systematic, and integrated luminosity uncertainties, respectively.
The uncertainties in $A\epsilon$ and in the FONLL predictions are the total uncertainties.
The last row (``Inclusive bin") presents the measured total cross section and the FONLL and \PYTHIA predictions for the phase-space region of $10 \le \pt^{\PB} < 17 \GeV$ and $\abs{y^{\PB}} < 1.45$, and
$17 \le \pt^{\PB} < 100 \GeV$ and  $\abs{y^{\PB}} < 2.1$.}
\def\arraystretch{1.25}
\newlength\ghmtabletwo
\ifthenelse{\boolean{cms@external}}{\setlength\ghmtabletwo{0.9\textwidth}}{\setlength\ghmtabletwo{\columnwidth}}
\resizebox{\ghmtabletwo}{!}{
\begin{tabular}[h]{ccccccc}
\hline
$\pt^{\PB}$ $[\GeVns{}]$ & $\abs{y^{\PB}}$ & $n_\text{sig}$ & $A\epsilon$ $[\%]$ & ${{\rd}\sigma}/{{\rd}\pt^{\PB}}$ $[\mub/\GeVns{}]$  & FONLL $[\mub/{\GeVns}]$ & \PYTHIA $[\mub/{\GeVns}]$ \\
\hline
10-13 & ${<}1.45$ & $408\,^{+52}_{-53}$ & $0.78\pm0.10$ & $3.0 \pm0.4 \pm0.4 \pm0.1$ & $1.4\,^{+0.5}_{-0.4}$ & $2.6$\\
13-17 & ${<}1.45$ & $755\,^{+47}_{-45}$ & $3.6\pm0.3$ & $0.88 \pm0.05 \pm0.08 \pm0.02$ & $0.62\,^{+0.21}_{-0.14}$ & $1.12$\\
17-24 & ${<}2.1$ & $1140\,^{+40}_{-39}$ & $7.1\pm0.6$ & $0.39 \pm 0.01 \pm0.04 \pm0.01$ & $0.30\,^{+0.08}_{-0.06}$ & $0.48$\\
24-30 & ${<}2.1$ & $519\,^{+30}_{-28}$ & $13\pm1\x$ & $0.12 \pm 0.01 \pm0.01 \pm0.00$ & $0.10 \pm {0.02}$ & $0.14$\\
30-40 & ${<}2.1$ & $404\,^{+24}_{-23}$ & $17\pm2\x$ & $(4.1 \pm 0.2 \pm0.4 \pm0.1) \times 10^{-2}$ & $(3.3\,^{+0.6}_{-0.5})\times 10^{-2}$ & $4.4\times 10^{-2}$\\
40-50 & ${<}2.1$ & $157\pm13$ & $20\pm2\x$ & $(1.3 \pm 0.1 \pm0.2 \pm0.0) \times 10^{-2}$ & $(1.0\,^{+0.2}_{-0.1})\times 10^{-2}$ & $1.3\times 10^{-2}$\\
50-60 & ${<}2.1$ & $49\pm8$ & $21\pm2\x$ & $(4.0\,^{+0.7}_{-0.6} \pm0.5 \pm0.1) \times 10^{-3}$ & $(3.9\,^{+0.6}_{-0.5})\times 10^{-3}$ & $4.5\times 10^{-3}$\\
60-70 & ${<}2.1$ & $23\,^{+6}_{-5}$ & $21\pm3\x$ & $(1.9\,^{+0.5}_{-0.4} \pm0.3 \pm0.0) \times 10^{-3}$ & $(1.7 \pm 0.2)\times 10^{-3}$ & $1.8\times 10^{-3}$\\
\x70-100 & ${<}2.1$ & $24\,^{+5}_{-4}$ & $20\pm3\x$ & $(6.7\,^{+1.4}_{-1.3} \pm1.0 \pm0.2) \times 10^{-4}$ & $(5.0 \pm 0.6)\times 10^{-4}$ & $5.0\times 10^{-4}$\\
\hline
$\pt^{\PB}$ $[\GeVns{}]$ & $\abs{y^{\PB}}$ & $n_\text{sig}$ & $A\epsilon$ $[\%]$ & ${{\rd}\sigma}/{{\rd}y^{\PB}}$ $[\mub]$  & FONLL $[\mub]$ & \PYTHIA $[\mub]$ \\
\hline
10-100 & 0.0--0.2 & $460\,^{+43}_{-33}$ & $3.6\pm0.5$ & $5.5\,^{+0.5}_{-0.4} \pm0.8 \pm0.1$ & $3.2\,^{+1.1}_{-0.7}$ & $5.7$\\
10-100 & 0.2--0.4 & $511\pm32$ & $3.8\pm0.5$ & $5.7\,^{+0.3}_{-0.4} \pm0.8 \pm0.1$ & $3.2\,^{+1.1}_{-0.7}$ & $5.7$\\
10-100 & 0.4--0.6 & $455\,^{+28}_{-27}$ & $4.0\pm0.5$ & $4.8 \pm0.3 \pm0.6 \pm0.1$ & $3.2\,^{+1.1}_{-0.7}$ & $5.6$\\
10-100 & \x0.6--0.85 & $576\,^{+30}_{-29}$ & $4.4\pm0.6$ & $4.5 \pm0.2 \pm0.6 \pm0.1$ & $3.1\,^{+1.1}_{-0.7}$ & $5.6$\\
10-100 & 0.85--1.1\x & $622\,^{+36}_{-35}$ & $4.2\pm0.6$ & $5.0 \pm0.3 \pm0.7 \pm0.1$ & $3.1\,^{+1.0}_{-0.7}$ & $5.4$\\
10-100 & \x1.1--1.45 & $671\,^{+42}_{-41}$ & $3.5\pm0.4$ & $4.6 \pm0.3 \pm0.6 \pm0.1$ & $2.9\,^{+1.0}_{-0.7}$ & $5.2$\\
17-100 & 1.45--1.8\x & $188\,^{+18}_{-17}$ & $4.4\pm0.4$ & $1.05\,^{+0.10}_{-0.09} \pm0.11 \pm0.02$ & $0.68\,^{+0.18}_{-0.13}$ & $1.05$\\
17-100 & 1.8--2.1 & $35\pm8$ & $1.4\pm0.2$ & $0.74\,^{+0.18}_{-0.16} \pm0.09 \pm0.02$ & $0.61\,^{+0.16}_{-0.12}$ & $0.96$\\
\hline
& & $n_\text{sig}$ & $A\epsilon$ $[\%]$ & ${\sigma}$ $[\mub]$  & FONLL $[\mub]$ & \PYTHIA $[\mub]$ \\
\hline
\multicolumn{2}{c}{Inclusive bin} & $3477\,^{+86}_{-84}$ & $3.9\pm0.5$ & $15.3 \pm0.4 \pm2.1 \pm0.4$ & $9.9\,^{+3.3}_{-2.2}$ & $17.2$\\
\hline
\end{tabular}
}
\label{tab:crosssection}
\end{table*}
\ifthenelse{\boolean{cms@external}}{}{\clearpage}
\section{Summary}
\label{sec:summary}
The differential cross sections for $\PBp$ meson production in $\Pp\Pp$ collisions at $\sqrt{s}=13\TeV$
have been measured for the first time by the CMS experiment using the decay channel $\PBp \to \PJGy \PKp$, with $\PJGy \to \mu^+\mu^-$,
as a function of $\pt^{\PB}$ for $\abs{y^{\PB}} < 1.45$ or  $\abs{y^{\PB}} < 2.1$, and as a function of $y^{\PB}$ for
$10 < \pt^{\PB} < 100\GeV$ or $17 < \pt^{\PB} < 100\GeV$.
The total cross section summed over all bins is measured to be $15.3 \pm0.4\stat\pm2.1\syst\pm0.4\lum\unit{\mub}$.
The measured distributions show reasonable agreement in terms of shape, as well as normalization,
with FONLL calculations and with the prediction of the \PYTHIA event generator, within the uncertainties.
The ratios between the measurements at 13 and at 7\TeV tend to prefer higher values compared to the predictions.
This study provides the first measurement of a b hadron cross section through the $\PBp \to \PJGy \PKp$ exclusive decay channel at the centre-of-mass energy of 13\TeV.

\begin{acknowledgments}
We congratulate our colleagues in the CERN accelerator departments for the excellent performance of the LHC and thank the technical and administrative staffs at CERN and at other CMS institutes for their contributions to the success of the CMS effort. In addition, we gratefully acknowledge the computing centres and personnel of the Worldwide LHC Computing Grid for delivering so effectively the computing infrastructure essential to our analyses. Finally, we acknowledge the enduring support for the construction and operation of the LHC and the CMS detector provided by the following funding agencies: BMWFW and FWF (Austria); FNRS and FWO (Belgium); CNPq, CAPES, FAPERJ, and FAPESP (Brazil); MES (Bulgaria); CERN; CAS, MoST, and NSFC (China); COLCIENCIAS (Colombia); MSES and CSF (Croatia); RPF (Cyprus); SENESCYT (Ecuador); MoER, ERC IUT and ERDF (Estonia); Academy of Finland, MEC, and HIP (Finland); CEA and CNRS/IN2P3 (France); BMBF, DFG, and HGF (Germany); GSRT (Greece); OTKA and NIH (Hungary); DAE and DST (India); IPM (Iran); SFI (Ireland); INFN (Italy); MSIP and NRF (Republic of Korea); LAS (Lithuania); MOE and UM (Malaysia); BUAP, CINVESTAV, CONACYT, LNS, SEP, and UASLP-FAI (Mexico); MBIE (New Zealand); PAEC (Pakistan); MSHE and NSC (Poland); FCT (Portugal); JINR (Dubna); MON, RosAtom, RAS and RFBR (Russia); MESTD (Serbia); SEIDI and CPAN (Spain); Swiss Funding Agencies (Switzerland); MST (Taipei); ThEPCenter, IPST, STAR and NSTDA (Thailand); TUBITAK and TAEK (Turkey); NASU and SFFR (Ukraine); STFC (United Kingdom); DOE and NSF (USA).
\end{acknowledgments}
\bibliography{auto_generated}

\providecommand{\href}[2]{#2}\begingroup\raggedright\begin{thebibliography}{10}%
\makeatletter
\providecommand{\hrefCMSnoop }[0]{\@secondoftwo}%
\makeatother
\providecommand{\doi}{\texttt{doi:}\begingroup \urlstyle{tt}\Url}

\bibitem{UA1-87}
\hrefCMSnoop {}{{UA1} Collaboration, ``{Beauty production at the CERN
  proton-antiproton collider}'',} \textit{ Phys. Lett. B} \textbf{ 186} (1987)
  237,
\href{http://dx.doi.org/10.1016/0370-2693(87)90287-5}{\doi{10.1016/0370-2693(87)90287-5}}.

\bibitem{UA188}
\hrefCMSnoop {}{{UA1} Collaboration, ``{Measurement of the bottom quark
  production cross section in proton-antiproton Collisions at $\sqrt{s} =
  0.63\TeV$}'',} \textit{ Phys. Lett. B} \textbf{ 213} (1988) 405,
\href{http://dx.doi.org/10.1016/0370-2693(88)91785-6}{\doi{10.1016/0370-2693(88)91785-6}}.

\bibitem{TeVI-CDF3}
\hrefCMSnoop {}{{CDF} Collaboration, ``{Measurement of the $\PB$ meson
  differential cross-section, $\rd \sigma / \rd \PT$, in $\Pp\bar{\Pp}$
  collisions at $\sqrt{s} = 1.8\TeV$}'',} \textit{ Phys. Rev. Lett.} \textbf{
  75} (1995) 1451,
  \href{http://dx.doi.org/10.1103/PhysRevLett.75.1451}{\doi{10.1103/PhysRevLett.75.1451}},
\href{http://www.arXiv.org/abs/hep-ex/9503013}{\texttt{arXiv:hep-ex/9503013}}.

\bibitem{TeVI-CDF6}
\hrefCMSnoop {}{{CDF} Collaboration, ``{Measurement of the $\PBp$ total cross
  section and $\PBp$ differential cross section $\rd \sigma/ \rd\PT$ in
  $\Pp\PAp$ collisions at $\sqrt{s}= 1.8\TeV$}'',} \textit{ Phys. Rev. D}
  \textbf{ 65} (2002) 052005,
  \href{http://dx.doi.org/10.1103/PhysRevD.65.052005}{\doi{10.1103/PhysRevD.65.052005}},
\href{http://www.arXiv.org/abs/hep-ph/0111359}{\texttt{arXiv:hep-ph/0111359}}.

\bibitem{TeVII-CDF1}
\hrefCMSnoop {}{{CDF} Collaboration, ``{Measurement of the $\PJGy$ meson and
  b-hadron production cross sections in $\Pp\PAp$ collisions at $\sqrt{s} =
  1960\GeV$}'',} \textit{ Phys. Rev. D} \textbf{ 71} (2005) 032001,
  \href{http://dx.doi.org/10.1103/PhysRevD.71.032001}{\doi{10.1103/PhysRevD.71.032001}},
\href{http://www.arXiv.org/abs/hep-ex/0412071}{\texttt{arXiv:hep-ex/0412071}}.

\bibitem{TeVII-CDF2}
\hrefCMSnoop {}{{CDF} Collaboration, ``{Measurement of the $\PBp$ production
  cross section in $\Pp\PAp$ collisions at $\sqrt{s} = 1960\GeV$}'',} \textit{
  Phys. Rev. D} \textbf{ 75} (2007) 012010,
  \href{http://dx.doi.org/10.1103/PhysRevD.75.012010}{\doi{10.1103/PhysRevD.75.012010}},
\href{http://www.arXiv.org/abs/hep-ex/0612015}{\texttt{arXiv:hep-ex/0612015}}.

\bibitem{TeVI-D01}
\hrefCMSnoop {}{{D0} Collaboration, ``{Inclusive $\Pgm$ and $\PQb$ quark
  production cross-sections in $\Pp\PAp$ collisions at $\sqrt{s} =
  1.8\TeV$}'',} \textit{ Phys. Rev. Lett.} \textbf{ 74} (1995) 3548,
\href{http://dx.doi.org/10.1103/PhysRevLett.74.3548}{\doi{10.1103/PhysRevLett.74.3548}}.

\bibitem{TeVI-D04}
\hrefCMSnoop {}{{D0} Collaboration, ``{Cross section for b jet production in
  $\PAp\Pp$ collisions at $\sqrt{s} = 1.8\TeV$}'',} \textit{ Phys. Rev. Lett.}
  \textbf{ 85} (2000) 5068,
  \href{http://dx.doi.org/10.1103/PhysRevLett.85.5068}{\doi{10.1103/PhysRevLett.85.5068}},
\href{http://www.arXiv.org/abs/hep-ex/0008021}{\texttt{arXiv:hep-ex/0008021}}.

\bibitem{Aad:2012jga}
\hrefCMSnoop {}{{ATLAS Collaboration}, ``{Measurement of the $\PQb$-hadron
  production cross section using decays to $\PD^{\ast}\PGmm {\rm X}$ final
  states in $\Pp\Pp$ collisions at $\sqrt{s} = 7\TeV $ with the ATLAS
  detector}'',} \textit{ Nucl. Phys. B} \textbf{ 864} (2012) 341,
  \href{http://dx.doi.org/10.1016/j.nuclphysb.2012.07.009}{\doi{10.1016/j.nuclphysb.2012.07.009}},
\href{http://www.arXiv.org/abs/1206.3122}{\texttt{arXiv:1206.3122}}.

\bibitem{ATLAS:2013cia}
\hrefCMSnoop {}{{ATLAS Collaboration}, ``{Measurement of the differential
  cross-section of $\PBp$ meson production in $\Pp\Pp$ collisions at $\sqrt{s}
  = 7\TeV $ at ATLAS}'',} \textit{ JHEP} \textbf{ 10} (2013) 042,
  \href{http://dx.doi.org/10.1007/JHEP10(2013)042}{\doi{10.1007/JHEP10(2013)042}},
\href{http://www.arXiv.org/abs/1307.0126}{\texttt{arXiv:1307.0126}}.

\bibitem{Khachatryan:2011mk}
\hrefCMSnoop {}{{CMS Collaboration}, ``{Measurement of the $\PBp$ production
  cross section in $\Pp\Pp$ collisions at $\sqrt{s} = 7\TeV $}'',} \textit{
  Phys. Rev. Lett.} \textbf{ 106} (2011) 112001,
  \href{http://dx.doi.org/10.1103/PhysRevLett.106.112001}{\doi{10.1103/PhysRevLett.106.112001}},
\href{http://www.arXiv.org/abs/1101.0131}{\texttt{arXiv:1101.0131}}.

\bibitem{Khachatryan:2011hf}
\hrefCMSnoop {}{{CMS Collaboration}, ``{Inclusive b-hadron production cross
  section with muons in $\Pp\Pp$ collisions at $\sqrt{s} = 7\TeV $}'',}
  \textit{ JHEP} \textbf{ 03} (2011) 090,
  \href{http://dx.doi.org/10.1007/JHEP03(2011)090}{\doi{10.1007/JHEP03(2011)090}},
\href{http://www.arXiv.org/abs/1101.3512}{\texttt{arXiv:1101.3512}}.

\bibitem{Chatrchyan:2011pw}
\hrefCMSnoop {}{{CMS Collaboration}, ``{Measurement of the $\PBz$ production
  cross section in $\Pp\Pp$ collisions at $\sqrt{s}=7\TeV $}'',} \textit{ Phys.
  Rev. Lett.} \textbf{ 106} (2011) 252001,
  \href{http://dx.doi.org/10.1103/PhysRevLett.106.252001}{\doi{10.1103/PhysRevLett.106.252001}},
\href{http://www.arXiv.org/abs/1104.2892}{\texttt{arXiv:1104.2892}}.

\bibitem{Chatrchyan:2011vh}
\hrefCMSnoop {}{{CMS Collaboration}, ``{Measurement of the $\PBzs$ production
  cross section with $\PBzs \to \PJGy \ensuremath{\phi}$ decays in $\Pp\Pp$
  collisions at $\sqrt{s}=7\TeV $}'',} \textit{ Phys. Rev. D} \textbf{ 84}
  (2011) 052008,
  \href{http://dx.doi.org/10.1103/PhysRevD.84.052008}{\doi{10.1103/PhysRevD.84.052008}},
\href{http://www.arXiv.org/abs/1106.4048}{\texttt{arXiv:1106.4048}}.

\bibitem{Chatrchyan:2012xg}
\hrefCMSnoop {}{{CMS Collaboration}, ``{Measurement of the $\PgL_{\rm b}$ cross
  section and the $\PagL_{\rm b}$ to $\PgL_{\rm b}$ ratio with $\PJGy \PgL$
  decays in $\Pp\Pp$ collisions at $\sqrt{s}=7\TeV $}'',} \textit{ Phys. Lett.
  B} \textbf{ 714} (2012) 136,
  \href{http://dx.doi.org/10.1016/j.physletb.2012.05.063}{\doi{10.1016/j.physletb.2012.05.063}},
\href{http://www.arXiv.org/abs/1205.0594}{\texttt{arXiv:1205.0594}}.

\bibitem{Chatrchyan:2012hw}
\hrefCMSnoop {}{{CMS Collaboration}, ``{Measurement of the cross section for
  production of $\PQb\PAQb {\rm X}$, decaying to muons in $\Pp\Pp$ collisions
  at $\sqrt{s}=7\TeV $}'',} \textit{ JHEP} \textbf{ 06} (2012) 110,
  \href{http://dx.doi.org/10.1007/JHEP06(2012)110}{\doi{10.1007/JHEP06(2012)110}},
\href{http://www.arXiv.org/abs/1203.3458}{\texttt{arXiv:1203.3458}}.

\bibitem{Khachatryan:2014nfa}
\hrefCMSnoop {}{{CMS Collaboration}, ``{Measurement of the ratio of the
  production cross sections times branching fractions of $\PB_{\PQc}^{\pm} \to
  \PJGy \Pgppm$ and $\PBpm \to \PJGy \PKpm $ and $\mathcal{B}(\PB_{\PQc}^{\pm}
  \to \PJGy
  \Pgppm\Pgppm\ensuremath{\pi^\mathrm{\mp}})/\mathcal{B}(\PB_{\PQc}^{\pm} \to
  \PJGy \Pgppm)$ in $\Pp\Pp$ collisions at $\sqrt{s} = 7 \TeV$}'',} \textit{
  JHEP} \textbf{ 01} (2015) 063,
  \href{http://dx.doi.org/10.1007/JHEP01(2015)063}{\doi{10.1007/JHEP01(2015)063}},
\href{http://www.arXiv.org/abs/1410.5729}{\texttt{arXiv:1410.5729}}.

\bibitem{Aaij:2012jd}
\hrefCMSnoop {}{{LHCb Collaboration}, ``{Measurement of the $\PBpm$ production
  cross-section in $\Pp\Pp$ collisions at $\sqrt{s}=7\TeV $}'',} \textit{ JHEP}
  \textbf{ 04} (2012) 093,
  \href{http://dx.doi.org/10.1007/JHEP04(2012)093}{\doi{10.1007/JHEP04(2012)093}},
\href{http://www.arXiv.org/abs/1202.4812}{\texttt{arXiv:1202.4812}}.

\bibitem{Aaij:2013noa}
\hrefCMSnoop {}{{LHCb Collaboration}, ``{Measurement of $\PB$ meson production
  cross-sections in proton-proton collisions at $\sqrt{s} = 7\TeV $}'',}
  \textit{ JHEP} \textbf{ 08} (2013) 117,
  \href{http://dx.doi.org/10.1007/JHEP08(2013)117}{\doi{10.1007/JHEP08(2013)117}},
\href{http://www.arXiv.org/abs/1306.3663}{\texttt{arXiv:1306.3663}}.

\bibitem{Aaij:2014ija}
\hrefCMSnoop {}{{LHCb Collaboration}, ``{Measurement of $\PB_{\PQc}^+$
  production in proton-proton collisions at $\sqrt{s}=8\TeV $}'',} \textit{
  Phys. Rev. Lett.} \textbf{ 114} (2015) 132001,
  \href{http://dx.doi.org/10.1103/PhysRevLett.114.132001}{\doi{10.1103/PhysRevLett.114.132001}},
\href{http://www.arXiv.org/abs/1411.2943}{\texttt{arXiv:1411.2943}}.

\bibitem{Aaij:2015rla}
\hrefCMSnoop {}{{LHCb Collaboration}, ``{Measurement of forward $\PJGy$
  production cross-sections in $\Pp\Pp$ collisions at $\sqrt{s}=13$\,\TeV}'',}
  \textit{ JHEP} \textbf{ 10} (2015) 172,
  \href{http://dx.doi.org/10.1007/JHEP10(2015)172}{\doi{10.1007/JHEP10(2015)172}},
\href{http://www.arXiv.org/abs/1509.00771}{\texttt{arXiv:1509.00771}}.

\bibitem{Aaij:2016avz}
\hrefCMSnoop {}{{LHCb Collaboration}, ``{Measurement of the $\PQb$-quark
  production cross-section in 7 and 13 TeV $\Pp\Pp$ collisions}'',} \textit{
  Phys. Rev. Lett.} \textbf{ 118} (2017) 052002,
  \href{http://dx.doi.org/10.1103/PhysRevLett.118.052002}{\doi{10.1103/PhysRevLett.118.052002}},
\href{http://www.arXiv.org/abs/1612.05140}{\texttt{arXiv:1612.05140}}.

\bibitem{Sjostrand:2007gs}
\hrefCMSnoop {}{T.~Sj{\"o}strand, S.~Mrenna, and P.~Z. Skands, ``{A brief
  introduction to PYTHIA 8.1}'',} \textit{ Comput. Phys. Commun.} \textbf{ 178}
  (2008) 852,
  \href{http://dx.doi.org/10.1016/j.cpc.2008.01.036}{\doi{10.1016/j.cpc.2008.01.036}},
\href{http://www.arXiv.org/abs/0710.3820}{\texttt{arXiv:0710.3820}}.

\bibitem{Cacciari:1998it}
\hrefCMSnoop {}{M.~Cacciari, M.~Greco, and P.~Nason, ``{The $\PT$ spectrum in
  heavy flavor hadroproduction}'',} \textit{ JHEP} \textbf{ 05} (1998) 007,
  \href{http://dx.doi.org/10.1088/1126-6708/1998/05/007}{\doi{10.1088/1126-6708/1998/05/007}},
\href{http://www.arXiv.org/abs/hep-ph/9803400}{\texttt{arXiv:hep-ph/9803400}}.

\bibitem{Cacciari:2001td}
\hrefCMSnoop {}{M.~Cacciari, S.~Frixione, and P.~Nason, ``{The $\pt$ spectrum
  in heavy flavor photoproduction}'',} \textit{ JHEP} \textbf{ 03} (2001) 006,
  \href{http://dx.doi.org/10.1088/1126-6708/2001/03/006}{\doi{10.1088/1126-6708/2001/03/006}},
\href{http://www.arXiv.org/abs/hep-ph/0102134}{\texttt{arXiv:hep-ph/0102134}}.

\bibitem{Chatrchyan:2012xi}
\hrefCMSnoop {}{{CMS Collaboration}, ``{Performance of CMS muon reconstruction
  in $\Pp\Pp$ collision events at $\sqrt{s} = 7\TeV $}'',} \textit{ JINST}
  \textbf{ 7} (2012) P10002,
  \href{http://dx.doi.org/10.1088/1748-0221/7/10/P10002}{\doi{10.1088/1748-0221/7/10/P10002}},
\href{http://www.arXiv.org/abs/1206.4071}{\texttt{arXiv:1206.4071}}.

\bibitem{:2008zzk}
\hrefCMSnoop {}{{CMS Collaboration}, ``The {CMS} experiment at the {CERN
  LHC}'',} \textit{ JINST} \textbf{ 3} (2008) S08004,
\href{http://dx.doi.org/10.1088/1748-0221/3/08/S08004}{\doi{10.1088/1748-0221/3/08/S08004}}.

\bibitem{Agashe:2014kda}
\hrefCMSnoop {}{{Particle Data Group}, K.~A. Olive {et~al.}, ``Review of
  particle physics'',} \textit{ Chin. Phys. C} \textbf{ 38} (2014) 090001 and
  2015 update,
  \href{http://dx.doi.org/10.1088/1674-1137/38/9/090001}{\doi{10.1088/1674-1137/38/9/090001}}.

\bibitem{Khachatryan:2015pea}
\hrefCMSnoop {}{{CMS Collaboration}, ``{Event generator tunes obtained from
  underlying event and multiparton scattering measurements}'',} \textit{ Eur.
  Phys. J. C} \textbf{ 76} (2016) 155,
  \href{http://dx.doi.org/10.1140/epjc/s10052-016-3988-x}{\doi{10.1140/epjc/s10052-016-3988-x}},
\href{http://www.arXiv.org/abs/1512.00815}{\texttt{arXiv:1512.00815}}.

\bibitem{Ball:2013hta}
\hrefCMSnoop {}{{NNPDF} Collaboration, ``{Parton distributions with QED
  corrections}'',} \textit{ Nucl. Phys. B} \textbf{ 877} (2013) 290,
  \href{http://dx.doi.org/10.1016/j.nuclphysb.2013.10.010}{\doi{10.1016/j.nuclphysb.2013.10.010}},
\href{http://www.arXiv.org/abs/1308.0598}{\texttt{arXiv:1308.0598}}.

\bibitem{GEANT4:2003}
\hrefCMSnoop {}{{GEANT4} Collaboration, ``{GEANT4---a simulation toolkit}'',}
  \textit{ Nucl. Instrum. Meth. A} \textbf{ 506} (2003) 250,
\href{http://dx.doi.org/10.1016/S0168-9002(03)01368-8}{\doi{10.1016/S0168-9002(03)01368-8}}.

\bibitem{Lange:2001}
\hrefCMSnoop {}{D.~J. Lange, ``{The \EVTGEN particle decay simulation
  package}'',} \textit{ Nucl. Instrum. Meth. A} \textbf{ 462} (2001) 152,
  \href{http://dx.doi.org/10.1016/S0168-9002(01)00089-4}{\doi{10.1016/S0168-9002(01)00089-4}}.

\bibitem{Khachatryan:2010yr}
\hrefCMSnoop {}{{CMS Collaboration}, ``{Prompt and non-prompt $\PJGy$
  production in $\Pp\Pp$ collisions at $\sqrt{s}=7$\,\TeV}'',} \textit{ Eur.
  Phys. J. C} \textbf{ 71} (2011) 1575,
  \href{http://dx.doi.org/10.1140/epjc/s10052-011-1575-8}{\doi{10.1140/epjc/s10052-011-1575-8}},
\href{http://www.arXiv.org/abs/1011.4193}{\texttt{arXiv:1011.4193}}.

\bibitem{TRK-11-001}
\hrefCMSnoop {}{{CMS Collaboration}, ``{Description and performance of track
  and primary-vertex reconstruction with the CMS tracker}'',} \textit{ JINST}
  \textbf{ 9} (2014) P10009,
  \href{http://dx.doi.org/10.1088/1748-0221/9/10/P10009}{\doi{10.1088/1748-0221/9/10/P10009}},
\href{http://www.arXiv.org/abs/1405.6569}{\texttt{arXiv:1405.6569}}.

\bibitem{LUM-15-001}
\href {https://cds.cern.ch/record/2138682}{{CMS Collaboration}, ``Cms
  luminosity measurement for the 2015 data taking period'',} CMS Physics
  Analysis Summary CMS-PAS-LUM-15-001, 2016.

\bibitem{Skwarnicki:1986xj}
\href {https://inspirehep.net/record/230779/}{T.~Skwarnicki, ``{A study of the
  radiative cascade transitions between the $\PgU^\prime$ and $\PgU$
  resonances}''}.
\newblock PhD thesis, Cracow, INP, 1986.
\newblock {DESY-F31-86-02}.

\bibitem{Ball:2014uwa}
\hrefCMSnoop {}{{NNPDF} Collaboration, ``{Parton distributions for the LHC Run
  II}'',} \textit{ JHEP} \textbf{ 04} (2015) 040,
  \href{http://dx.doi.org/10.1007/JHEP04(2015)040}{\doi{10.1007/JHEP04(2015)040}},
\href{http://www.arXiv.org/abs/1410.8849}{\texttt{arXiv:1410.8849}}.

\bibitem{Botje:2011sn}
M.~Botje\hrefCMSnoop {}{ {et~al.}, ``{The PDF4LHC Working Group interim
  recommendations}'',} (2011).
\href{http://www.arXiv.org/abs/1101.0538}{\texttt{arXiv:1101.0538}}.

\bibitem{Alekhin:2011sk}
\hrefCMSnoop {}{S.~Alekhin {et~al.}, ``{The PDF4LHC Working Group interim
  report}'',} (2011).
\href{http://www.arXiv.org/abs/1101.0536}{\texttt{arXiv:1101.0536}}.

\end{thebibliography}\endgroup
\cleardoublepage \appendix\section{The CMS Collaboration \label{app:collab}}\begin{sloppypar}\hyphenpenalty=5000\widowpenalty=500\clubpenalty=5000\textbf{Yerevan Physics Institute,  Yerevan,  Armenia}\\*[0pt]
V.~Khachatryan, A.M.~Sirunyan, A.~Tumasyan
\vskip\cmsinstskip
\textbf{Institut f\"{u}r Hochenergiephysik der OeAW,  Wien,  Austria}\\*[0pt]
W.~Adam, E.~Asilar, T.~Bergauer, J.~Brandstetter, E.~Brondolin, M.~Dragicevic, J.~Er\"{o}, M.~Flechl, M.~Friedl, R.~Fr\"{u}hwirth\cmsAuthorMark{1}, V.M.~Ghete, C.~Hartl, N.~H\"{o}rmann, J.~Hrubec, M.~Jeitler\cmsAuthorMark{1}, A.~K\"{o}nig, I.~Kr\"{a}tschmer, D.~Liko, T.~Matsushita, I.~Mikulec, D.~Rabady, N.~Rad, B.~Rahbaran, H.~Rohringer, J.~Schieck\cmsAuthorMark{1}, J.~Strauss, W.~Treberer-Treberspurg, W.~Waltenberger, C.-E.~Wulz\cmsAuthorMark{1}
% ...
\vskip\cmsinstskip
% ...
1:~~Also at Vienna University of Technology, Vienna, Austria\\
% ...

\end{sloppypar}
\end{document}